%% file: jet_eff.tex
\documentclass[fleqn,usenatbib]{mnras}
\usepackage{amsmath}
\usepackage{graphicx}
\usepackage{amssymb,txfonts}
\usepackage[usenames, dvipsnames]{color} 

\newcommand{\fermi}{{\textit{Fermi}}}
\newcommand {\ea} {{et~al.}}

\newcommand{\g}{$\gamma$}

\DeclareRobustCommand{\VAN}[3]{#2} 

\title{The power and production efficiency of blazar jets}

\author[Pjanka \ea]{Patryk Pjanka,$^{1}$ Andrzej A.~Zdziarski$^{2}$ and Marek Sikora$^{2}$\\ 
$^{1}$Department of Astrophysical Sciences, Princeton University, 4 Ivy Lane, Princeton, NJ 08544, USA \\
$^{2}$Nicolaus Copernicus Astronomical Center, Polish Academy of Sciences, Bartycka 18, 00-716 Warsaw, Poland 
}

\pagerange{\pageref{firstpage}--\pageref{lastpage}}
\pubyear{2016}

\begin{document} 

\maketitle
\label{firstpage}

\begin{abstract}
We use published data on the power and production efficiency of jets in blazars with double radio lobes in order to compare results obtained using different methods. In order to eliminate selection effects, we use cross-matched sub-samples containing only luminous blazars. We compare the three main existing methods, namely those based on the emission of radio lobes, on spectral fitting, and on radio core shift. We find the average jet power obtained for identical samples with the radio-lobe method is $\sim$10 times lower than that from the spectral fitting. In turn, the power from spectral fitting is compatible with that from core-shift modelling for plausible parameters of the latter. We also consider a phenomenological estimator based on the \g-ray luminosity. We examine uncertainties of those methods and discuss two alternative hypotheses. In one, the blazar-fit and core-shift methods are assumed to be correct, and the lower power from radio lobes is caused by intermittency of accretion. Long periods of quiescence cause the energy in the radio lobes, accumulated over the lifetime of the blazar, to be much less than that estimated based on the present luminous state. In addition, the power calculated using the radio lobes can be underestimated for intrinsically compact jets, in which the radio core flux can be over-subtracted. In our second hypothesis, the radio-lobe method is assumed to be correct, and the blazar-fit and core-shift powers are reduced due to the presence of $\sim$15 pairs per proton and a larger magnetization than usually assumed, respectively.
\end{abstract}
\begin{keywords}
galaxies:~active -- galaxies:~jets -- quasars:~general -- radiation~mechanisms:~non-thermal -- gamma-rays:~galaxies
\end{keywords}

\section{Introduction}\label{introduction}

Starting with the seminal paper by \citet{Rawlings1991}, it has become common to use energetics and lifetimes of extended double radio sources to calculate jet powers in radio galaxies and quasars. Such studies indicated that in radio-loudest objects the jet power, $P_{\rm j}$, is comparable or even exceeds the accretion power, $\dot M c^2$, where $\dot M$ is the accretion rate (see, e.g., \citealt{Punsly2007, Fernandes2011, Sikora2013}). Problems with launching such powerful jets by standard accretion discs stimulated studies on the so-called magnetically arrested discs (MAD; \citealt{Narayan2003}). In scenarios involving the MAD, the jet is powered by a fast spinning black hole (BH) immersed in a strong magnetic field supported by the ram pressure of the accretion flow \citep{Tchekhovskoy2011, McKinney2012}. As it has been theoretically estimated and numerically confirmed, the spin-extraction/MAD scenario allows launching jets with the power up to $\simeq\! 3 \dot M c^2$ or so. 

The applicability of the spin-extraction/MAD model to radio loud quasars was questioned by \citet{vanVelzen2013}, who, by using energetics of radio lobes, found that the median value of the jet production efficiency, $\eta_{\rm j} \equiv P_{\rm j}/\dot M c^2$, in a radio-selected sample of quasars with double radio sources is $\sim$0.01. If $\eta_{\rm j}$ depended only on the BH spin, this would imply its average value much lower than those predicted by BH cosmological evolution \citep{Volonteri2013}.  
 
On the other hand, large efficiencies, with $\eta_{\rm j}\sim 0.1$--10, have been found using blazar models (\citealt{Ghisellini14}, hereafter G14). Similar efficiencies were then obtained by means of core-shift measurements in radio-core dominated quasars (\citealt{Zamaninasab14}, hereafter Z14; \citealt{Zdziarski15}, hereafter Z15). In this paper, we compare results obtained by all of these methods. In order to minimize effects of selection biases in comparison of different samples, we use cross-matched sub-samples.

The paper is organized as follows. We define our observational samples in Section~\ref{samples}. Section~\ref{methods} presents the methods used to calculate the jet and accretion powers. Section~\ref{results} presents our results, which are then discussed in Section~\ref{discussion}.
 
\section{Samples}\label{samples}

We use published samples of blazars containing information on their various properties. These include broad-band spectra, radio core-shift measurements, luminosities of extended radio structures, and accretion luminosities. Some of those samples, in addition to luminous blazars (i.e., radio-loud quasars seen close to their jet axes), include BL Lac objects and radio galaxies. Since accretion flows in most of those sources are probably radiatively inefficient, it is difficult to measure their accretion rates needed to estimate $\eta_{\rm j}$.  Hence, we study only quasar-associated blazars, also called flat-spectrum radio quasars (FSRQs). Those objects have the morphology compatible with those of FR-II sources, as indicated by their extended radio structure and large radio luminosity. Each sample is denoted by the first letter of the first-author name. They are as follows.

\begin{itemize}

\item \citet{Kharb2010} (K) used the Very Large Baseline Array (VLBA) MOJAVE-I sample \citep{Lister2009}, for which they measured the extended radio flux by subtracting core emission from the VLA images. We use these subtracted values. As can be seen in their Fig.~4, majority of the sources of \cite{Kharb2010} have radio luminosity at 1400MHz higher than $10^{25}\textrm{ W}/\textrm{Hz}$, which, in view of the correlations between radio luminosity and morphology of radio structures of \cite{Ledlow1996} and \cite{Best2009}, confirms the FRII character of those sources.

\item \citet{Meyer2011} (M) obtained a sample of quasars and BL Lac objects chosen from multiple catalogues with broad-band spectra and/or core-subtracted images from the Very Large Array (VLA). We use their values of the extended and core-subtracted 300-MHz luminosity for quasars. \cite{Meyer2011} do not report $1400$MHz luminosities of their objects and, hence, it is difficult to directly confirm their FRII morphology. However, the agreement between our results for the K and M samples lets us conclude that cases of FRI sources are also here negligibly rare.

\item G14 (G) studied 191 quasars with detections by the \fermi\/ Large Area Telescope (LAT), known broad-line region (BLR) luminosity, and a sufficient broad-band coverage. For those objects, they calculated the jet power and the bulk Lorentz factor, $\Gamma$, using blazar-model spectral fitting. They also calculated the accretion luminosity from both spectral fitting of a disc model, $L_{\rm d,fit}$, and by assuming it is $L_{\rm d,BLR}=$ 10 times the luminosity of the BLR, where the BLR luminosity was estimated based on the H$\beta$, Mg{\sc ii} and C{\sc iv} spectral lines. We use all of these values.

\item \citet{Arshakian2012} (A) used those sources from the MOJAVE-I sample that both appear in the 1$^{\textrm{st}}$ \fermi/LAT Catalogue \citep{1FGL} and have known optical nuclear fluxes and redshifts. This yields a sample of 76 blazars, for which we use their \g-ray fluxes. In our work, we also need the apparent jet velocities of these blazars, $\beta_{\rm app}$. We use the maximal values of $\beta_{\rm app}$ from \citet{Lister2009}, where they are based on observations of apparent movement of radio structures beyond radio cores. Hereafter, we assume the implied Lorentz factor to equal $\Gamma_{\rm app}=(1+\beta_{\rm app}^2)^{1/2}$, which is the minimum possible $\Gamma$ for a given $\beta_{\rm app}$. This relation is achieved for the viewing angle of $i\simeq 1/\Gamma$, which we also assume.

\item \citet{Pushkarev2012} (P) obtained values of radio core shift for a sub-sample of MOJAVE blazars for frequency pairs of 15.4--8.1, 15.4-8.4 and 15.4--12.1 GHz. For a given object, we use the weighted average of the angular shift divided by the wavelength difference. 

\item Z14 (Z) studied 59 blazars (including 54 FSRQs) chosen based on the availability of their redshift, $\Gamma_{\rm app}$, radio core shift (averaged from \citealt{Pushkarev2012}) and BH mass. They determined the accretion luminosity based on fluxes of the H$\beta$, Mg{\sc ii} and O{\sc iii} spectral lines using their eqs.\ (9--11), which method differs in details from the analogous BLR method of G14 (see also Section~\ref{results}). The sample of Z14 was later studied by Z15, who determined the jet power for each object based on the magnetic flux calculated using radio core shifts. We use their values of $\Gamma$ and the accretion luminosity, and calculate the core-shift values using weighted averages of the results for three pairs of frequencies of \citet{Pushkarev2012}. 

\item \citet{XiongZhang2014} (X) studied over 20 published blazar data sets combined with the 2$^{\textrm{nd}}$ \fermi/LAT Catalogue \citep{2FGL}. We use only their values of the accretion luminosity, $L_{\rm d,BLR}$, obtained using methods similar to that of G14, assuming it is 10 times the luminosity of the BLR.

\end{itemize}

In order to mitigate the selection effects present in comparison of samples of different origin, we use mostly sub-samples obtained by cross-matching. For instance, sources appearing in both \citet{Meyer2011} and G14 form a sample called M+G. In some cases, we also use sub-samples discriminating against sources from one of the main samples, e.g., G-M denotes the sources from G14 that are not present in \citet{Meyer2011}. For source cross-identification, we require both the location of sources within $1\degr$ of each other (corresponding approximately to the resolution of a 70-m radio telescope at 300 MHz), and the relative redshift difference $<10$ per cent. This is done, in particular, for the radio/\g-ray cross-matching.

\section{Methods and notation}
\label{methods}

\begin{table*}
 \centering
 \caption{Description of the main symbols. Here $L_{\rm d}$ is the accretion luminosity, $\Gamma$ is jet bulk Lorentz factor, and $P$ (omitting the subscript `j') is the jet power; their subscripts denote the method used to calculate them. The accretion rate is assumed to satisfy $\dot M c^2=10 L_{\rm d}$. Then symbols may have a superscript corresponding to the data set used.}
 \begin{tabular}{|c|l|l|c}
 \hline
 Symbol & Meaning and method of calculation & Data set & Sample/Superscript \\
 \hline
 $L_{\rm d,fit}$ & accretion luminosity from disc fitting & G14 &G\\
 $L_{\rm d,BLR}$ & accretion luminosity from BLR luminosity  & G14 &G\\
 & & \citet{XiongZhang2014} &X\\
 $L_{\rm d,line}$ & accretion luminosity from spectral lines & Z14 &Z\\
 $\Gamma_{\rm fit}$ & Lorentz factor from disc fitting & G14 &G\\
 $\Gamma_{\rm app}$ & Lorentz factor $=(1+\beta_{\rm app}^2)^{1/2}$ & \citet{Arshakian2012} &A\\
 && Z14 & Z\\
 $\Gamma_{\rm av}$ & average Lorentz factor $=13$ & G14 &G\\
 $P_{\rm fit}$ & jet power from blazar-model spectral fitting & G14 & G\\
 $P_{\rm cs}$ & jet power from radio core shift,  & \citet{Pushkarev2012} &P\\
& \ \ equations (7) and (25) of Z15  & Z14 & Z\\
 $P_{\rm rl}$ & jet power from radio lobes,  & \citet{Kharb2010} & K\\
& \ \ equation (11) of \citet{Willott1999}& \citet{Meyer2011} & M \\
 $P_\gamma$ & jet power from $\gamma$-ray flux, equation (\ref{Pgamma}) & \citet{Arshakian2012} & A\\
 \hline
 \label{symbols}
 \end{tabular}
\end{table*}

\begin{table*}
 \centering
 \caption{Default values and plausible parameter ranges of the model parameters. The given default values of $n_{\rm e}/n_{\rm p}$ and $\sigma_B$ correspond to the model without e$^\pm$ pairs. See Section~\ref{pairs} for their values in the model with pairs, which are approximately equal to their maximum values given here. 
}
 \begin{tabular}{|c|l|c|c}
 \hline
 Symbol & Meaning& Default value & Plausible range \\
 \hline
$n_{\rm e}/n_{\rm p}$ & Ratio of the e$^-+$\,e$^+$ density to that of protons &1 &1--30\\ 
$\beta_{\rm eq}$ & Ratio of the comoving particle kinetic energy density to $B^2/8\upi$ & 1 & 0.1--10\\
$\sigma_B$ & ratio of the magnetic to matter enthalpy fluxes & 0.03 &0.01--2\\
$\Theta\Gamma$ & observationally determined product of the opening angle and $\Gamma$ & 0.11 &0.1--0.2\\
$s$ & the causality parameter & 0.6 &0.5--1\\
$K_\gamma$ & the fudge factor of $P_\gamma$ & 2 &2--4\\
$f$ & the fudge factor of $P_{\rm rl}$ & 10 & 5--20\\
$\epsilon_{\rm d}$ & the accretion radiative efficiency & 0.1 & 0.05--0.3\\
 \hline
 \label{uncer}
 \end{tabular}
\end{table*}

We use three main methods to calculate the jet power. In addition, we use a fourth simple estimator, which can be normalized to agree on average with any of the other methods. They are listed in Table \ref{symbols}, which also summarizes the notation we use. Then Table~\ref{uncer} gives the default values and plausible parameter ranges of the model parameters.

The first method is that used by G14, see also \citet{Ghisellini09}. They use a one-zone blazar model to fit the broad-band spectra of blazars assuming one proton per electron; we denote the resulting power as $P_{\rm fit}$. The fits also provide the values of $\Gamma$, with the average of $\Gamma_{\rm av}\simeq 13$. Since these values are given by the fits together with the values of the jet power, we cannot quantify the effect of changing $\Gamma$ on the power in this method. We point out here that the jet power equals the enthalpy flux rather than that of energy, which follows from the form of the relativistic stress-energy tensor \citep{Levinson06}. Thus, the internal energy fluxes, used by G14, need to be multiplied by the corresponding adiabatic indices, which equal 4/3 for both highly relativistic plasma and tangled magnetic field, and 2 for toroidal magnetic field (e.g., \citealt{leahy91}). Thus, we multiply their components of the power in relativistic electrons by 4/3. For magnetic field, we assume equal parts of the toroidal and tangled (e.g., \citealt{Heinz00}) fields, which implies the magnetic adiabatic index to be $\kappa_B=5/3$. Given that the powers in G14 are dominated by cold ions, this is only a minor correction, unless there is a substantial amount of e$^\pm$ pairs, which then reduces the kinetic power in ions by the ratio of $n_{\rm e}\equiv n_- +n_+$ to that in protons, $n_{\rm p}$, and the total power in a more complex way, see Section~\ref{pairs}. G14 assumed $n_{\rm e}/n_{\rm p}=1$, which is the main systematic uncertainty of their results. Obviously, there is also some systematic uncertainty of their fitting method, which is difficult to estimate, and thus we do not attempt to do it here.

The second method is based on measuring the angular shift of the position of radio cores \citep*{Lobanov98,Shabala12}. The method assumes the conical jet model of \citet{Blandford79}, and yields the product of the  magnetic field, $B$, and the distance from the BH, $h$, see equation (7) of Z15 (which corrects a minor error in previous results), where $Bh$ is constant along the jet for toroidal field with conserved energy flux, and it weakly depends on the jet opening angle, $\Theta$, and the plasma beta (equipartition) parameter, $\beta_{\rm eq}$, as $\propto (\beta_{\rm eq}\Theta)^{-2/(6+p)}$, where $p$ is the electron power-law index. For $i\simeq 1/\Gamma$ (assumed by both Z14 and Z15), $Bh$ is also almost independent of $\Gamma$. The magnetic power then $P_B=(B h \Theta\Gamma )^2\beta c/2$, where $\beta c$ is the jet velocity. Z15 assumed $\Theta\Gamma =0.11$ based on the comparison of their two core-shift methods, and in agreement with the observational results showing that $\Theta\Gamma \simeq$0.1--0.2 on average \citep{Pushkarev09,Clausen13}. Treating $\Theta\Gamma$ as a fixed parameter, we have $P_B\propto [\Gamma/(\beta_{\rm eq}\Theta\Gamma)]^{4/(6+p)}$.

To calculate the total power, we need to relate the magnetic and matter enthalpy fluxes, which ratio is given by the magnetization parameter, $\sigma_B\simeq P_B/(P_{\rm j}-P_B)$, implying
\begin{equation}
P_{\rm j}=P_{\rm cs}=P_B(1+1/\sigma_B).
\label{pb}
\end{equation}
If we allow the uncertainties of $\sigma_B$ and $\beta_{\rm eq}$ as given in Table~\ref{uncer}, equation (\ref{pb}) implies a rather large uncertainty of $P_{\rm j}$. However, except for Section~\ref{pairs} where we consider a pair-dominated jet, we follow here Z15, who used the relation of
\begin{equation}
\sigma_B\simeq (\Theta\Gamma/s)^2,
\label{sigma}
\end{equation}
where the parameter $s$ is constrained by causality to be $\lesssim 1$ \citep{Tchekhovskoy2009}. For $\Theta\Gamma=0.11$, this yields $\sigma_B\simeq 0.012/s^2$. The total jet power is then $\propto s^2 [\Gamma/(\beta_{\rm eq}\Theta\Gamma)]^{4/(6+p)}$ and only weakly dependent on $\sigma_B$ as long as it is $<1$, 
\begin{equation}
P_{\rm j}=P_{\rm cs}=(B h s)^2 \beta c(1+\sigma_B)/2,
\label{pz15}
\end{equation}
similarly to equation (25) of Z15. We can then constrain the value of $s$ by imposing $\sigma_B$ to equal its (geometric) average value in the fits to the sample of G14, which is $\simeq 0.03$. This yields $s\simeq 0.6$, a plausible value, which we adopt hereafter (note that Z15 used $s=1$ in their figures). Then, the uncertainties in $s$, $\beta_{\rm eq}$ and $\Theta\Gamma$ given in Table~\ref{uncer} imply the uncertainty range of a factor of $\sim$0.3--10 of our obtained values of $P_{\rm j}$. 

The third method uses the emission of extended radio lobes and is based on the description of \citet{Willott1999}. Specifically, we use their equation (11), where $P_{\rm j}\propto f^{3/2}$, and $f$ is a fudge factor constrained to $f\lesssim 20$ \citep{Blundell00}. Here, we assume $f=10$. Additionally, we multiply their formula by 4/3 to account for the pressure work required to form the lobes (analogous to the enthalpy correction discussed above).  This method was critically reviewed by \citet{Godfrey13}, who pointed out that it did not include radiative energy losses of the radio lobes and that their inclusion introduces a significant dependence of the jet power on the size of the source (see \citealt{Shabala13}). They verified the dependence of the jet power on the radio luminosity by using data on hot spots associated with jet terminal shocks. Unfortunately, the hot spots are not visible or are not prominent enough in most FR-II sources to be used for measuring the jet powers. However, as we can see in fig.\ 3 of \citet{Godfrey13}, the jet powers from the hot-spot calorimetry for luminous FR-II sources agree well with the jet powers obtained using the \citet{Willott1999} method with $f=10$,  with the scatter in $P_j$ within a factor $\lesssim$2. We note that radio lobes are assumed by \cite{Willott1999} to be in equipartition, which minimizes their energy content (\citealt{1956Burbidge}, \citealt{1970Pacholczyk}; see also \citealt{2014Zdziarski}). However, minor departures from equipartition lead only to relatively minor increases of the jet power, e.g., by 1.15 for the magnetic field strength at 0.7 of the equipartition value (as in \citealt{2005Croston}).

Finally, we consider a purely phenomenological method based only on the \g-ray luminosity, $L_\gamma$. This is motivated by the \g-ray luminosity giving a major contribution to the bolometric luminosity, and the claims of \citet{Nemmen12} and G14 that the jet power is about 10 times higher than the radiative power (which assumes the validity of the first method with $n_{\rm e}=n_{\rm p}$). This then gives an estimate of the instantaneous jet power. We can then write (see equation 1 of G14),
\begin{equation}
P_{\rm j}=P_\gamma\simeq 2\times 10\frac{4}{3} \frac{K_\gamma L_\gamma}{\Gamma^2},
\label{Pgamma}
\end{equation}
where $L_\gamma$ is calculated for a single jet from \fermi/LAT data assuming isotropy, $K_\gamma$ is a fudge factor taking into account the bolometric correction and departures from the above value of 10 for the ratio of the jet power to the radiative power, and the factors of 2 and 4/3 follow from summing the contributions from the jet and counterjet and from considering Compton scattering of external radiation \citep{Ghisellini10}, respectively. We see that $P_{\rm j}\propto\Gamma^{-2}$ in this method. Initially, we assume $K_\gamma=2$, but we can adjust this coefficient to achieve for agreement with the results of any of the above methods.

We then estimate the jet production efficiency, $P_{\rm j}/\dot M c^2$. We assume the accretion radiative efficiency of $\epsilon_{\rm d}=0.1$, i.e., $\dot M c^2= L_{\rm d}/\epsilon_{\rm d}$, and use the three methods of estimating $L_{\rm d}$, one from disc fitting and two based on the BLR emission, see Section~\ref{samples} and Table~\ref{symbols}. 

\section{Results}\label{results}

\renewcommand{\arraystretch}{1.2}
\setlength{\tabcolsep}{2pt}
\begin{table}
 \centering
 \caption{The number of sources, $N$, logarithmic averages, standard deviations, $\sigma$, and medians for selected quantities in the studied samples. The parameter $s=0.6$ is assumed for the core-shift method.}
 \begin{tabular}{cccccc}
 \hline
\input{Table2.tex}
 \hline
 \label{averages}
 \end{tabular}
\end{table}

\begin{figure}
 \centering
 \includegraphics[width=\columnwidth]{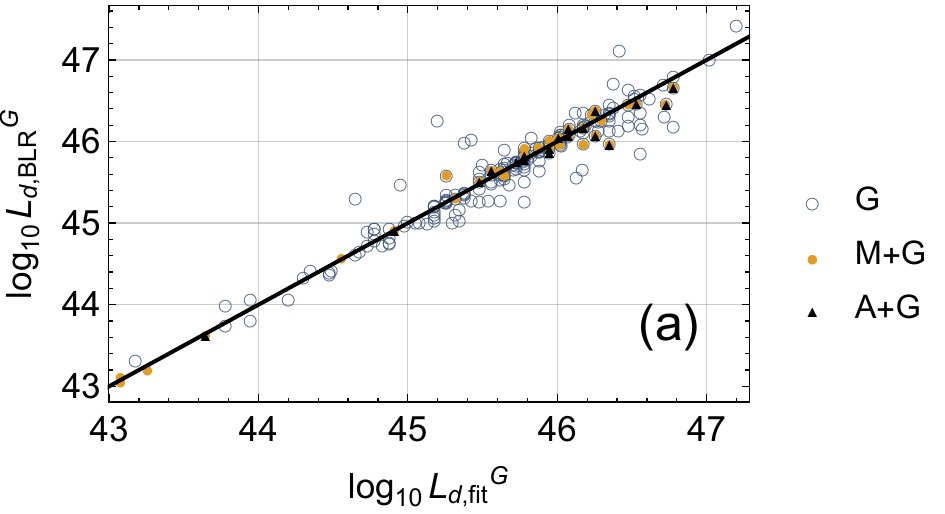}
 \includegraphics[width=\columnwidth]{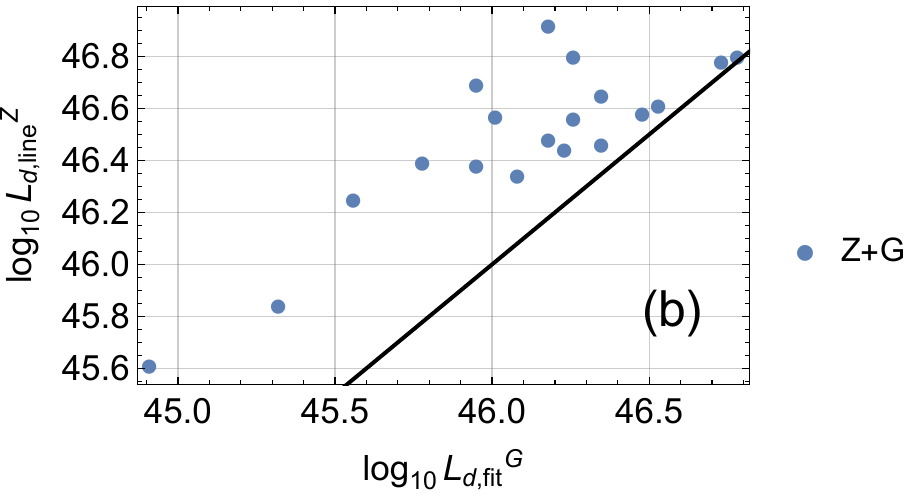}
 \includegraphics[width=\columnwidth]{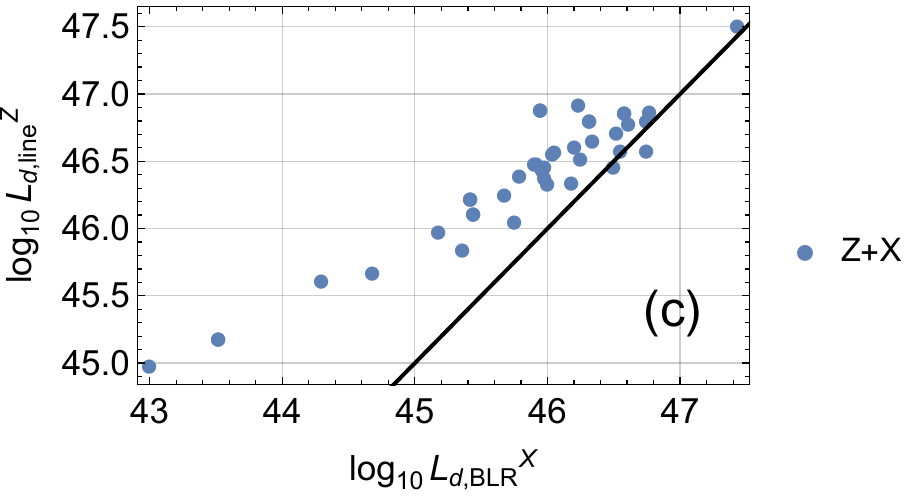}
 \caption{The comparison of accretion luminosities calculated using different methods and cross-matched samples, as specified on each panel. 
}
 \label{lacc}
 \end{figure}

\begin{figure}
 \centering
 \includegraphics[width=\columnwidth]{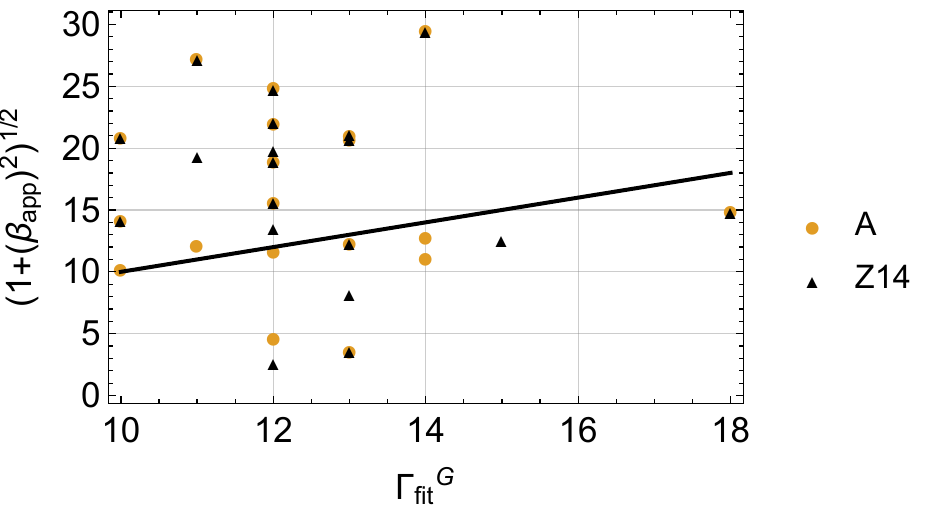}
 \caption{The comparison of bulk Lorentz factors calculated using the fitting method of G14 with those based on the apparent velocities from \citet{Lister2009} for the samples of \citet{Arshakian2012} and Z14.
}
 \label{gamma}
 \end{figure}

Before proceeding to the issue of the jet power, we compare different methods of estimating the accretion luminosity, see Fig.~\ref{lacc} and Table~\ref{averages}. We find very good agreement, on average, between the spectral fit method of G14 and their estimates based on the BLR lines, see Fig.~\ref{lacc}(a). Similarly, we find good agreement between either method of G14 and the BLR method of \citet{XiongZhang2014}, see Table~\ref{averages}. However, we find a rather substantial disagreement between the values of $L_{\rm d}$ obtained using the method of Z14 and all of the other three methods, namely the BLR and fit methods of G14 and the BLR method of \citet{XiongZhang2014}, see Table~\ref{averages}. We illustrate two of those discrepancies in Figs.\ \ref{lacc}(b--c). We see the values obtained by Z14 are higher than the other ones by factors $\sim$2.5--3 on average. 

This appears to result from different methods applied to obtain accretion luminosities in these cases. While \citet{XiongZhang2014} and G14 use the fits of \citet{Calderone2013} and \citet{Celotti1997} to relate the luminosities of broad spectral lines to the bolometric accretion disc luminosity, Z14 obtains their accretion luminosities using the set of relations from \citet{PunslyZhang2011}. In both of them the accretion luminosity can be expressed as a power law of each of the line luminosity, $L_{\rm acc} \propto L_{\rm line}^{\alpha}$, but they differ in the value of $\alpha$. For example, $\alpha = 0.78$ is used for the H$\beta$ line in \citet{Punsly2011}, while \citet{Calderone2013} uses a linear relation. This difference appears to explain the systematic differences seen in Fig.~\ref{lacc}(b--c). An answer to the question which of the accretion luminosity estimators is more robust requires analyses beyond the scope of this paper. However, the aforementioned systematic differences need to be borne in mind during analysis of the jet production efficiencies.

Both our sample of \citet{Arshakian2012} and Z14 use the maximal apparent velocities reported in \citet{Lister2009}, so the relation of $\Gamma_{\rm app}$ to $\Gamma_{\rm fit}^{\rm G}$ is similar in both those cases, see Fig.~\ref{gamma}. However, we find that the Lorentz factors resulting from apparent motions are higher on average than those from blazar model fitting. Namely, we find $\Gamma_{\rm app}>\Gamma_{\rm fit}$ for 12/19 and 13/19 in the samples of \citet{Arshakian2012} and Z14, respectively. On average, the values of $\Gamma_{\rm app}$ are higher than of $\Gamma_{\rm fit}$ by $\simeq$4. We discuss this issue in Section~\ref{discussion}.

We then compare the jet power estimations, see Fig.~\ref{power} and Table~\ref{averages}. Those numbers bear substantial uncertainties following, among other, from the uncertainties of the model parameters, see Table~\ref{uncer}. However, these uncertainties are difficult to assess quantitatively, and we do not give them in Table~\ref{averages}. Instead, we discuss those uncertainties in Section~\ref{uncertainties}. 

We find some highly significant systematic differences between the jet powers obtained by different methods. The jet powers from radio lobes, $P_{\rm rl}$, are on average a factor of $\sim$10 lower than those from blazar model fitting, $P_{\rm fit}$, see Fig.~\ref{power}(a). At the value of its parameter of $s=0.6$, the jet powers obtained using radio core shifts, $P_{\rm cs}$, are on average almost the same as those from the blazar model fitting, $P_{\rm fit}$, see Fig.~\ref{power}(b). Thus, the average ratio between the powers from the core-shift and radio-lobe methods is also $\sim$10, as illustrated in Fig.~\ref{power}(c). Finally, the phenomenological values based on the \g-ray luminosity are, on average, lower by a factor of $\sim$2 than the corresponding values of $P_{\rm fit}$, see Fig.~\ref{power}(d), as well as of $P_{\rm cs}$. This may be due to an underestimate by a factor of $\sim$2 of $K_\gamma$, the constant in equation (\ref{Pgamma}). 

Finally, we compare the jet production efficiencies found by the different methods, see Fig.~\ref{hist} and Table~\ref{averages}. We present the results as plots of the probability density function (PDF), i.e., the integral of each histogram is normalized to unity. We show a few histograms on a panel, with each of them normalized separately. Each bar in a histogram begins at zero. Table~\ref{averages} gives the average and median values of $\log_{10}\eta_{\rm j}$. Naturally, the results here follow from those on $P_{\rm j}$ and $L_{\rm d}$. We see that the lowest values of $\eta_{\rm j}$ are found from the radio-lobe method, with both their average and median $<0.1$. The values from blazar fitting and core shift are much higher, with the average efficiencies of $\simeq$1. The values from the core-shift method and using the disc luminosities either from the BLR method or the fit of G14 are similar to those of the fitting method, with the average efficiencies of $\simeq$1. Using the core-shift method together with disc luminosities obtained using the spectral line method of Z14 again gives a lower average, $\simeq$0.3, which is due to the higher disc luminosities, as discussed above, see Fig.~\ref{lacc}(b--c). 

\begin{figure}
 \centering
 \includegraphics[width=7.4cm]{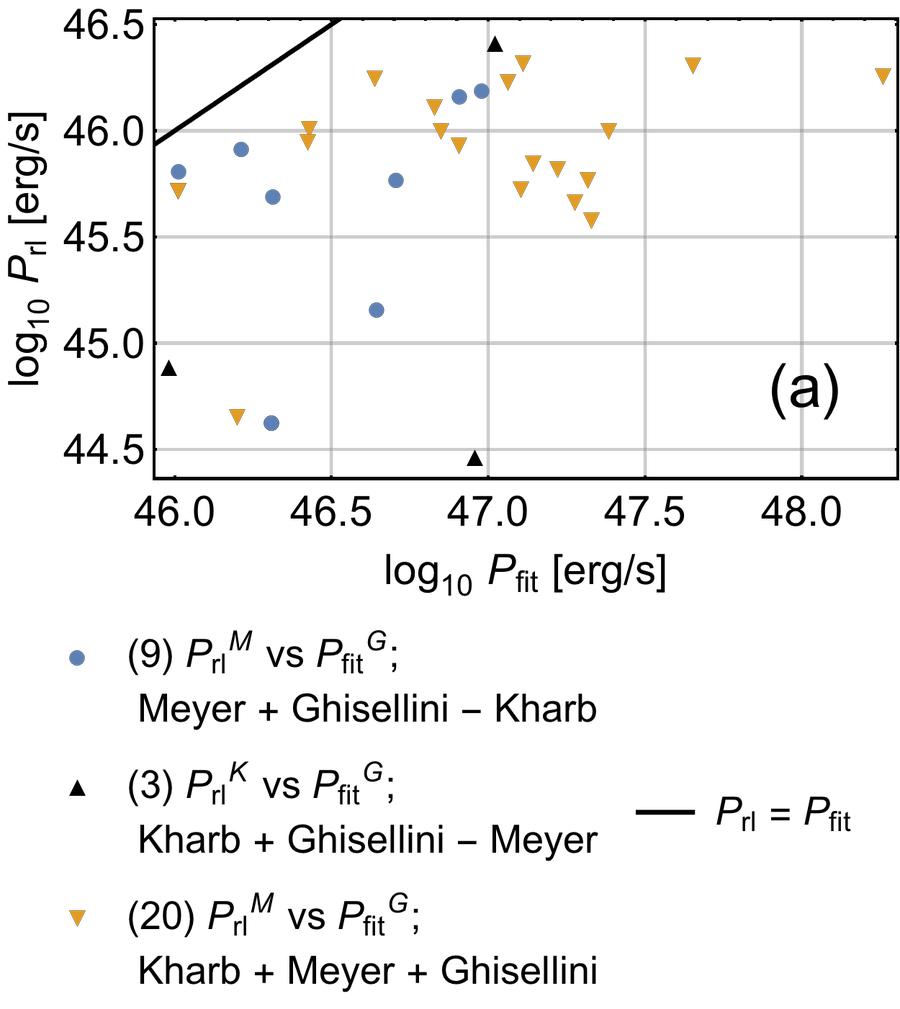}
 \includegraphics[width=7.4cm]{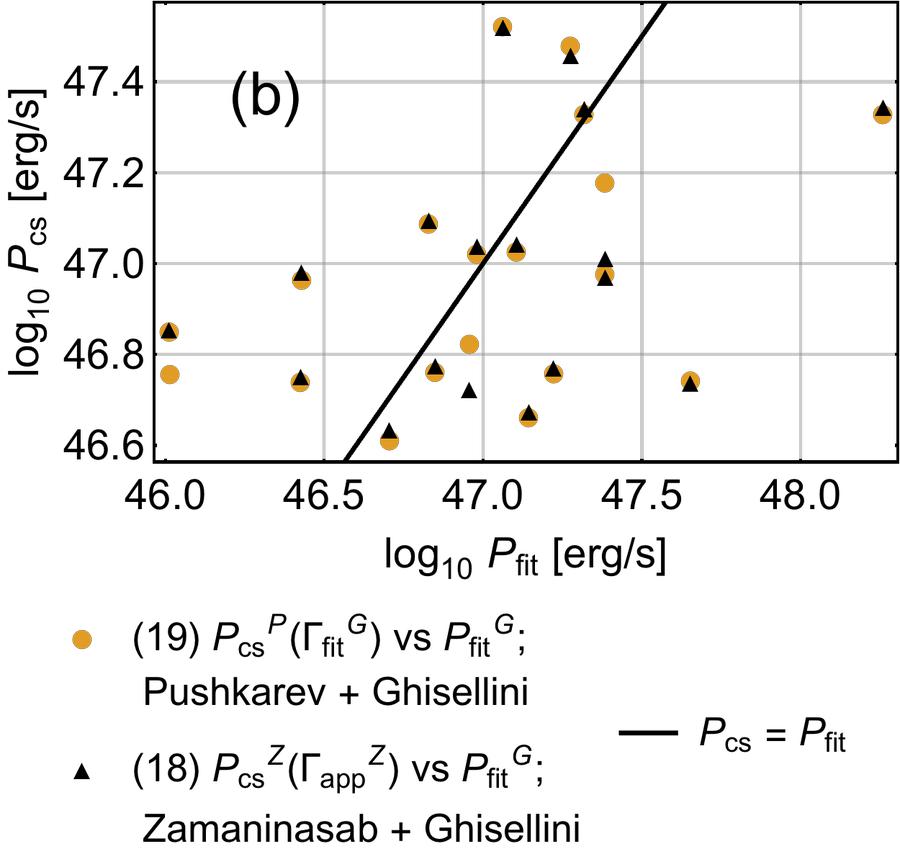}
 \includegraphics[width=7.4cm]{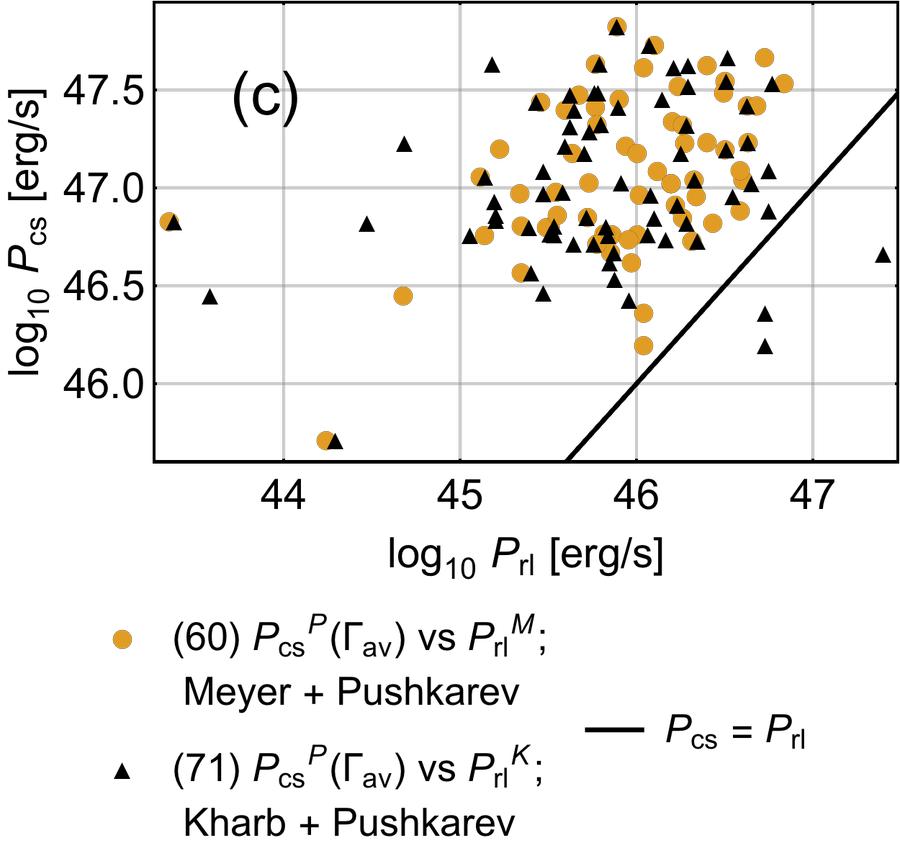}
 \caption{The comparison of jet powers calculated using different methods and cross-matched samples, as specified below the panels. The numbers in parentheses give the number of sources in each sample.
}
 \label{power}
 \end{figure}
 
\setcounter{figure}{2}
\begin{figure}
 \centering
 \includegraphics[width=7.4cm]{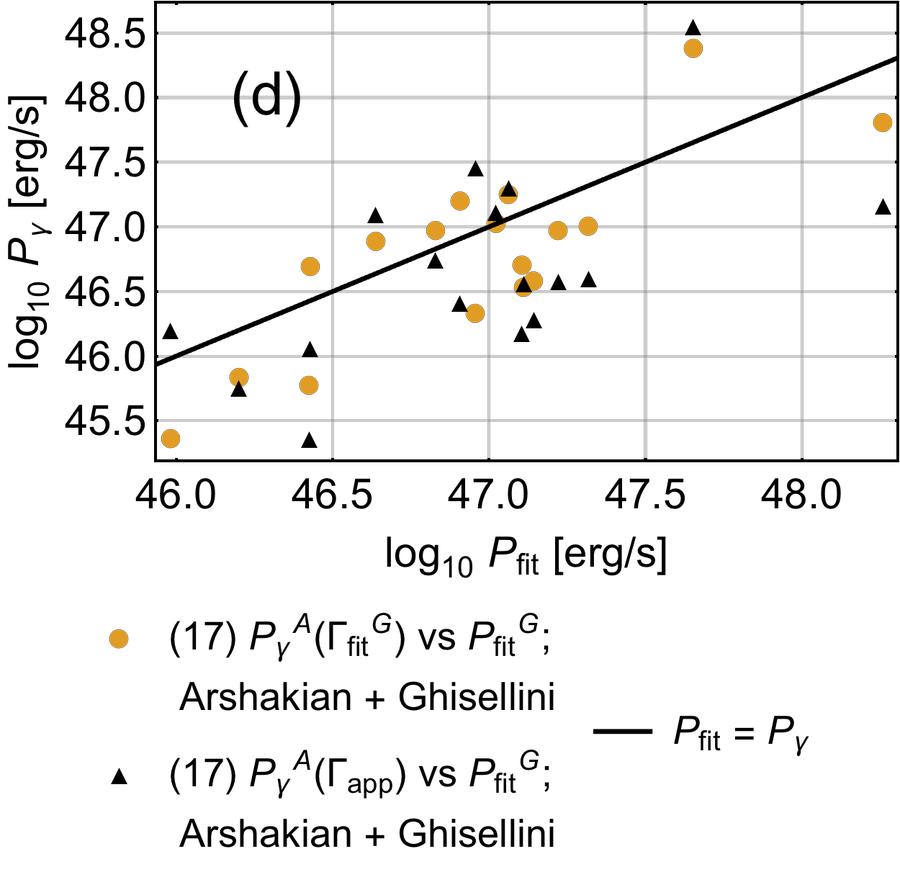}
 \caption{Continued.}
\end{figure}

\begin{figure}
 \centering
 \includegraphics[width=6.9cm]{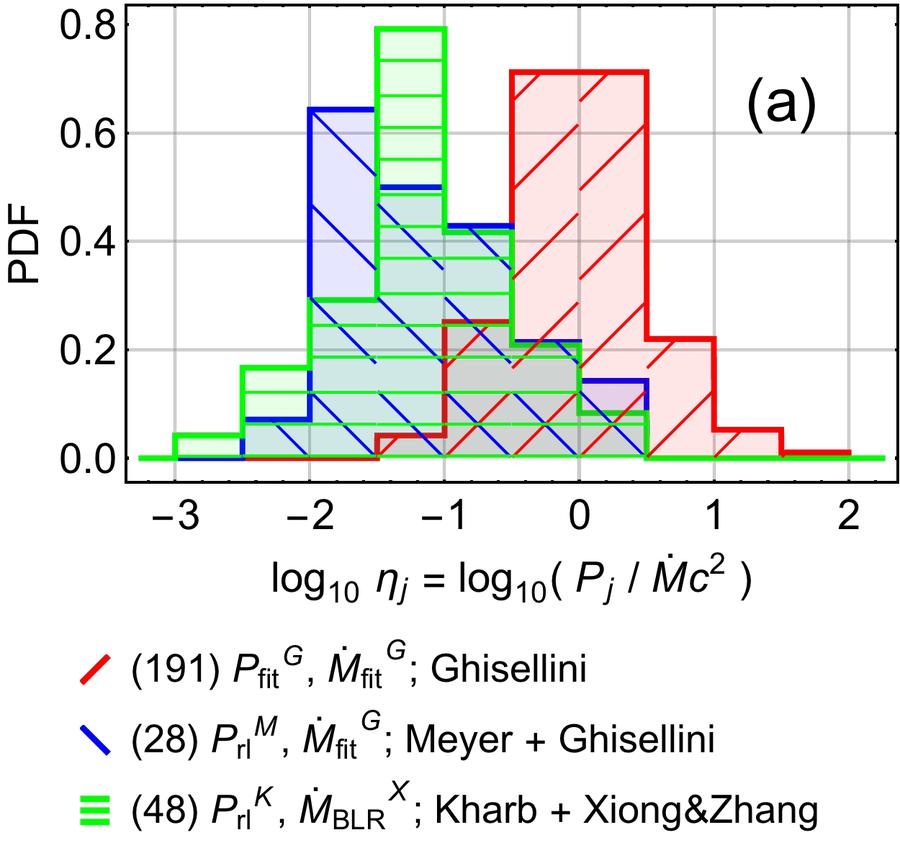}
 \includegraphics[width=6.9cm]{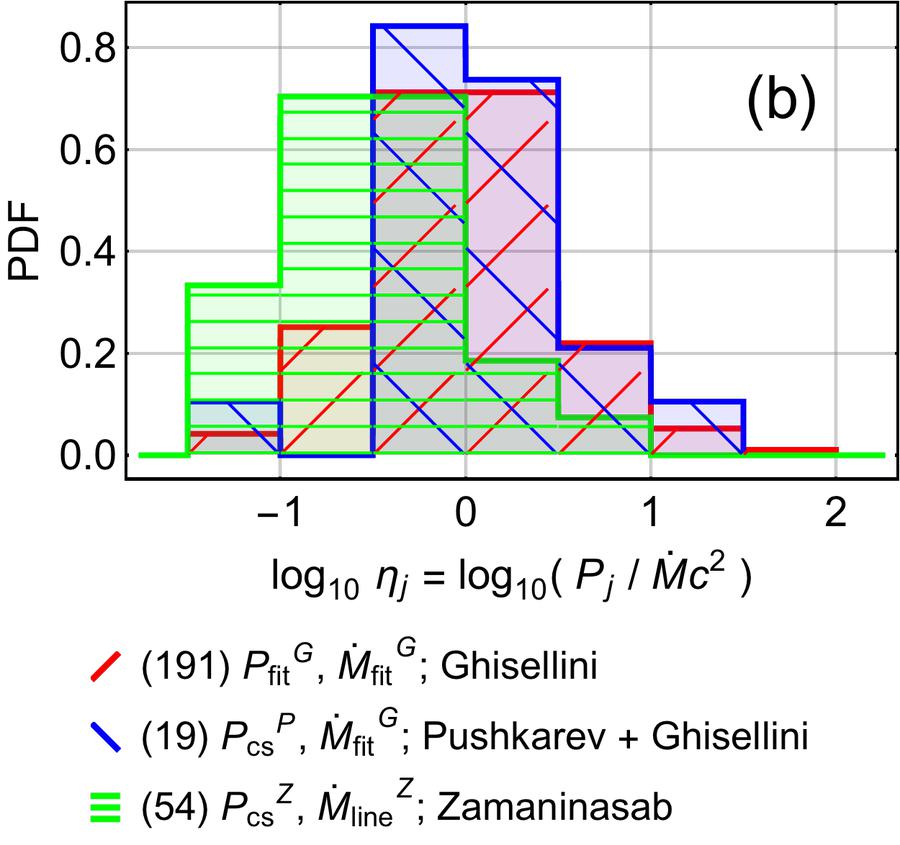}
 \caption{The comparison of distributions of the jet production efficiency calculated using different methods, described below each histogram. The numbers in parentheses give the number of sources in each sample.  }
 \label{hist}
 \end{figure}

\section{Discussion}\label{discussion}

\subsection{Uncertainties of the models}
\label{uncertainties}

We first discuss the discrepancies found in the estimates of jet power. 
They are related in part to the systematic uncertainties in the methods, which we have discussed in Section~\ref{methods}. 

The blazar-model powers, $P_{\rm fit}$, are sensitive only to the electron distribution in the jet, and assume one proton per electron, with the dominant component of the power due to cold protons (G14). If there is some amount of e$^\pm$ pairs, the jet power can be significantly reduced. We discuss this issue in detail below in Section~\ref{pairs}. Furthermore, the assumption of a one-zone uniform emission region is an oversimplification, with transversely stratified jets (proposed, e.g., by \citealt{Ghisellini05}) usually requiring lower power to reproduce the observed spectra \citep{Sikora2016}. 

The same issue will also affect the determination of the fitted jet Lorentz factor, $\Gamma_{\rm fit}$, in the blazar model (G14). If the jet is composed of a spine and a sheath moving at different Lorentz factors, the one-zone fit may yield inaccurate results, which may possibly explain the average $\Gamma_{\rm app} > \Gamma_{\rm fit}$ seen in Fig.~\ref{gamma}. On the other hand, the apparent jet speeds $\beta_{\rm app}$ used in these samples are the maximal values observed by \citet{Lister2009}. They may thus not correspond to the average jet velocity. Also, some of the observed $\beta_{\rm app}$ may trace a shock propagating through the jet, which we would expect to overtake the matter travelling with the relativistic flow. The values of $\Gamma$ affect the power from the core-shift and \g-ray luminosity methods in the way given in Section~\ref{methods}. 

The core-shift method makes two major assumptions. The first one is the adoption of the conical jet model of \citet{Blandford79}, with $B\propto h^{-1}$ and the density of relativistic electrons $\propto h^{-2}$ assumed in the core-shift calculations of \citet{Lobanov98}, \citet{Shabala12}, Z14 and Z15, which implies the radio spectral index in the partially synchrotron self-absorbed part of the spectrum of 0. This model is is clearly not strictly valid in the considered sources, as indicated by the radio spectral indices being $\neq 0$ in many blazars. This will lead to an inaccurate estimate of the actual magnetic field. Second, the core-shift method estimates the magnetic field times distance, $B h$, in the radio core region, and the resulting $P_B$ depends on $\Gamma$ and $\beta_{\rm eq}\Theta\Gamma$, see Section~\ref{methods}. The total power depends then both on those quantities and on the unknown value of the magnetization parameter, $\sigma_B$, see equation (\ref{pb}). We related, following Z15, $\sigma_B$ to $\Theta \Gamma/s$, and calibrated the results to those of G14 (see Section~\ref{methods}). This results in $P_{\rm j}$ being almost independent of $\sigma_B$, see equation (\ref{pz15}). However, the above relation is satisfied only if the jet acceleration and collimation are dominated by ideal MHD processes. While this is likely for $\sigma_B\gtrsim 1$, conversion of the Poynting flux to the kinetic energy leading to $\sigma_B \ll 1$, assumed here, is instead likely to be driven by dissipative processes (see Section~6 in Z15 and references therein). If $\sigma_B$ has a larger value than that assumed here, then $P_{\rm j}$ will be given by equation (\ref{pb}), and it will be lower, see Section~\ref{pairs}. Still, we find the average jet power in our samples from the core-shift and blazar-model methods are almost equal to each other for $\beta_{\rm eq}=1$ and our assumed value of the causality parameter, $s=0.6$, which followed from assuming $\sigma_B=0.03$, the average value from the blazar-model fits in G14, see Section~\ref{methods}.

We have seen that the main discrepancy is the jet power from radio lobes being much lower than those from the spectral fitting and from the core shifts. We first consider whether the low values of $P_{\rm rl}$ can be due to jet energy losses during propagation. Given that $P_{\rm rl}\ll P_{\rm fit},\,P_{\rm cs}$, this would require the loss of most of the jet power, which is inconsistent with the result of G14 that the radiative power is only $\sim$0.1 of the jet power. Then, a non-radiative process would, in principle, have to be called forth to explain such a reduction of the jet power. 

The radio-lobe power is $\propto f^{3/2}$, and the factor $f$ may, in principle, be twice the value of 10 assumed here. Still, this will not bring the results of this method to agreement with the jet powers from the other methods. However, an additional factor may, in principle, arise from some of the radio-lobe luminosities being underestimated due to the assumption that blazars can exist only in quasars with classical double radio sources, i.e., those characterized by sizes larger than tens of kpc. Contrary to this assumption, some blazars can be hosted by younger or shorter-living radio sources, as represented by compact symmetric objects and GHz-peaked sources. Then, the methods of spectral decomposition and core subtraction used by \citet{Meyer2011} and \citet{Kharb2010} to calculate radio-lobe luminosities would lead to underestimation of the lobe flux, and thus the jet power, for those objects. However, only $\sim$7 per cent of blazars do not exhibit extended radio-emission \citep{Kharb2010}, so this effect appears unlikely to explain the observed systematic differences between $P_{\rm rl}$ and the other estimates of $P_{\rm j}$.

A possible explanation of this discrepancy is intermittency of accretion. Our samples are selected to consist of the most luminous objects, FSRQs, which have the Eddington ratios $\gtrsim 0.03$. It is likely that such high luminosity occurs only during a small fraction of the lifetime of a blazar. The methods based on the current state of the jet derive then the power averaged over relatively short, recent, time. On the other hand, the energy in the lobes has accumulated over the blazar lifetime, which can consist mostly of periods of quiescence. Thus, the ratio of $P_{\rm rl}$ to $P_{\rm fit}$, $P_{\rm cs}$ is equal to the ratio of the luminous-to-average powers. 

This also agrees with the results of \citet{vanVelzen2013}, who selected $\sim\! 10^5$ double radio sources from the Faint Images of the Radio Sky at Twenty-Centimeters (FIRST) radio catalogue, but found quasar counterparts for only $\sim\! 10^3$ sources. This is consistent with the sources being most of the time in quiescence.

The average uncertainty of the (phenomenological) method based on the \g-ray luminosity lies in the value of the fudge factor $K_\gamma$. If we assume the correctness of the method of G14, we obtain $K_\gamma\simeq 4$, but we can adjust this coefficient to achieve agreement with any of the other two methods. The \g-ray luminosity method is also the coarsest of the four, as it is based only on one quantity, which, e.g., has a strong dependence on the viewing angle due to the relativistic beaming. Thus, it may give large departures from the actual power in individual cases.

\subsection{The effect of e$\mathbf{^\pm}$ pairs}
\label{pairs}

On the other hand, it is still possible that there is a certain amount of pairs in the jet. Then, the jet power in the ion rest mass will be reduced by $n_{\rm p}/n_{\rm e}$. Following \citet{sikora16b}, we consider here consequences of the assumption that the radio-lobe determinations of the jet power are correct, and the presence of pairs reduces the powers from the blazar model to the values at which the average power ratio between the two methods is unity. The average values of the components of the jet and accretion power in G14 are: the power in ions assuming fully ionized H composition, i.e., a pure proton-electron plasma, which we denote here as $P_{\rm p=e}$, of $\simeq 2.5\times 10^{46}$ erg s$^{-1}$, the power in relativistic electrons, $P_{\rm e}\simeq 2.5\times 10^{44}$ erg s$^{-1}$, the power in magnetic field, $P_B\simeq 1.0\kappa_B \times 10^{45}$ erg s$^{-1}$, the radiative power, $P_{\rm rad}\simeq 2.0\times 10^{45}$ erg s$^{-1}$, and $\dot M c^2\simeq 3.2\times 10^{46}/(\epsilon_{\rm d}/0.1)$ erg s$^{-1}$. As in Section~\ref{methods}, we assume here $\kappa_B=5/3$. Then, $P_B\simeq 7 P_{\rm e}$, and we hereafter can neglect the contribution from $P_{\rm e}$.

Assuming that the power of the jet upon leaving the blazar zone, $P_{\rm j}=P_{\rm p}+P_{\rm e}+P_B$ (where $P_{\rm p}$ is the power in ions without any assumption about the pair content), equals the estimate from radio lobes, we obtain the jet power in ions, $P_{\rm p}$, of (neglecting the small contribution from $P_{\rm e}$)
\begin{equation}
P_{\rm p}=P_{\rm rl}-P_B.
\end{equation}
Since $P_{\rm p}=(n_{\rm p}/n_{\rm e})P_{\rm p=e}$,
\begin{equation}
\frac{n_{\rm e}}{n_{\rm p}}\simeq \frac{P_{\rm p=e}}{P_{\rm rl}-P_B}.
\label{pair_ratio}
\end{equation}
Using our estimate of $P_{\rm j}=P_{\rm rl}\simeq 0.1 P_{\rm fit}\sim 2.5\times 10^{45}$ erg s$^{-1}$ on average (Table~\ref{averages}) and $P_{\rm fit}\simeq P_{\rm p=e}$, we have $n_{\rm e}/n_{\rm p}\sim 30$, or $\sim$15 pairs per proton. We stress that this number is rather approximate, given the number of assumptions involved.

We then consider the total power supplied to the jet before any radiative losses, $P_{\rm j0}$. Then, again assuming the correctness of the radio-lobe power estimate, we have
\begin{equation}
P_{\rm j0}=P_{\rm rl}+P_{\rm rad}\sim 4.5\times 10^{45}\,{\rm erg}\,{\rm s}^{-1}.
\label{total}
\end{equation}
Thus, $\sim$1/2 of the power supplied to the jet is radiated, i.e., the jet radiative efficiency is $\sim$1/2. We can also estimate the magnetization parameter, $\sigma_B\simeq P_B/(P_{\rm j}-P_B)$, to be $\sigma_B\sim 2$ in the region of the blazar zone. This differs from the average in G14 of $\simeq$0.03 because of their assumption of no pair content.

\citet{sikora16b} also gives further arguments supporting a substantial pair content in jets. He points out that allowing for a substantial pair content allows for a large ratio of the external Compton to synchrotron powers, which is in agreement with observations, and gives the average electron Lorentz factor of $\sim$100, which then yields correct average energies of the synchrotron and external Compton peaks. 

We then consider the estimate from the core shift. As we pointed out in Sections \ref{methods} and \ref{uncertainties}, our method of determining $P_{\rm j}$ depends sensitively on the assumption connecting $\sigma_B$ with $\Theta\Gamma$. If we allow $\sigma_B\sim 1/2$ in the radio cores while keeping the assumed $\Theta\Gamma\simeq 0.11$ unchanged, we will reduce the average of $P_{\rm cs}$ to the average of $P_{\rm rl}$, see equation (\ref{pb}). This then favours magnetic field reconnection as the process powering the blazar radiation.

\subsection{The jet production efficiency}
\label{efficiency}

The jet production efficiency is an important criterion discriminating between various jet ejection models. While powers of jets launched by accretion discs are expected to be lower than the accretion powers, they can be comparable to or even slightly exceed the accretion power in the spin-extraction/MAD scenario. Hence, having observationally confirmed efficiencies $\eta_{\rm j} \gtrsim 1$ would strongly favour powering of jets by rotating BHs. Still, this efficiency is limited to $\lesssim 3$ \citep{McKinney2012}. This constraint comes from the fact that the magnetic flux confined on the BH by ram pressure of the accretion flow is limited by the accretion rate. Thus, any finding of $\eta_{\rm j} \gtrsim 3$, such as in a number of cases here, has to be due to either measurement errors, incorrect modelling or another jet ejection process being in place.

The jet production efficiency depends on the estimates of both the jet and accretion powers. We have compared the values of the latter obtained with the BLR and disc fitting methods, and found an overall good agreement between all those methods, see Table~\ref{averages} and Fig.~\ref{lacc}(a). On the other hand, we have found the accretion luminosities obtained by Z14 using a method based on selected spectral lines to give systematically higher values than the other approaches, by a factor of $\sim$3, see discussion in Section~\ref{results}. Then, the values of $\dot M c^2$ depend on the radiative efficiency, which we assume here to be 0.1. This is both uncertain and varying from one source to another. However, this uncertainty affects the obtained values of $\eta_{\rm j}$, but not the comparison of the results of different methods. Still, this might be responsible for the presence of the tail of sources with $\eta_{\rm j}>3$, due to an underestimation of their $\dot M$. 

We point out that large $\eta_{\rm j}$ requires both large initial spins and geometrically thick accretion flows \citep{Avara2016}. Spread of the evolutionarily determined spin \citep{Volonteri2013} and of the geometrical thickness are likely to be responsible for a broad range of $\eta_{\rm j}$ found by \citet{vanVelzen2013} in their sample. Their median value of $\eta_{\rm j}$ is lower by a factor of a few than that found here by the radio-lobe method, which is most likely a selection effect. Namely, \citet{vanVelzen2013} used a sample selected from the FIRST radio catalogue, which has much better flux sensitivity than the Cambridge catalogues, on which our samples are based. In particular, the values of the Eddington ratio for their sources, $\lesssim 0.1$, are lower than ours.

Finally, we stress that the presence of pairs will reduce the jet production efficiency. Given the value of $P_{\rm j0}$ in equation (\ref{total}), we have $\eta_{\rm j}=P_{\rm j0}/\dot M c^2\sim 0.15$. This removes the problem of having a substantial fraction of sources with $\eta_{\rm j}>3$.

The MAD scenario predicts $\eta_{\rm j}\sim a^2 (H/R)^q$, where $H/R$ is the disc height-to-radius ratio, $a$ is the BH dimensionless spin, and \citet{Avara2016} suggested $q\simeq 2$. This would imply $H/R\sim 1/3$, still much larger than those of the standard discs, but predicted by a number of theoretical studies, see, e.g., \citet{Begelman15}.

\section{Conclusions}

Our main conclusions are as follows. 

We have found the average jet power obtained for identical samples with the radio-lobe method is $\sim$10 times lower than that from the spectral fitting, under the assumption of no e$^\pm$ pairs. Then, the power from spectral fitting is similar that from core-shift modelling assuming the magnetization parameter in the radio cores is linked to the jet opening angle. We have also considered a phenomenological estimator based on the \g-ray luminosity, which can be adjusted to equal either of the methods by adjusting its normalization.

In order to resolve the discrepancy between radio-lobe jet powers and those from other methods, we have proposed two alternative hypotheses. First, the blazar-fit and core-shift methods are assumed to be correct, and the lower power from radio lobes is caused by intermittency of accretion, including long periods of quiescence, which causes the energy in the radio lobes, accumulated over the lifetime of the blazar, to be much less than that estimated based on the present luminous state. In addition, the power calculated using the radio lobes can be underestimated for intrinsically compact jets, in which the radio core flux can be over-subtracted. 

Alternatively, we have assumed that the radio-lobe method yields correct results. The blazar-fit method can be brought to agreement with it if there is certain amount of e$^\pm$ pairs, since the proton kinetic energy in radiative leptonic models is $\propto (n_{\rm p}/n_{\rm e})P_{\rm p=e}$. We find agreement between the two methods if there are $\sim$15 pairs per proton. Then the core-shift method can be brought to agreement with the radio-lobe one if the jet magnetization in the radio cores is larger than that implied by the jet opening angle, which can be the case if the jet acceleration and collimation is not driven by ideal MHD processes.

We cannot definitely decide which hypothesis is correct. It may also be that both discussed effects operate. The first would reduce the discrepancy between the radio-lobe and blazar-fit results, which would in turn reduce the estimated number of pairs per proton.

\section*{Acknowledgements}

We thank Tuomas Savolainen for valuable discussions, and the referee, Stanislav Shabala, for valuable suggestions and comments. We also thank Kenji Toma, who inspired us to work on the power of jets in blazars by finding that the powers computed using blazar models are substantially larger than those found using radio-lobe calorimetry (unpublished results, 2012). This research has been supported in part by the Polish National Science Centre grants 2012/04/M/ST9/00780, 2013/10/M/ST9/00729 and 2015/18/A/ST9/00746. 


 \bibliographystyle{mnras}
 \bibliography{references}

\label{lastpage}

\end{document}

%% file: Table2.tex
Sample & N & Quantity[$x$] & $\langle\log_{10}(x)\rangle$ & $\sigma[\log_{10}(x)]$ & $\log_{10}[{\rm median}(x)]$ \\
\hline
G & 191 & $L_{\rm d,fit}^{\rm G} / L_{\rm d,BLR}^{\rm G}$ & 0.0 & 0.1 & 0.0 \\
A+X+G & 17 & $L_{\rm d,fit}^{\rm G} / L_{\rm d,BLR}^{\rm X}$ & 0.0 & 0.1 & 0.0 \\
A+X+G & 17 & $L_{\rm d,BLR}^{\rm G} / L_{\rm d,BLR}^{\rm X}$ & 0.0 & 0.1 & 0.0 \\
Z+G & 18 & $L_{\rm d,fit}^{\rm G} / L_{\rm d,line}^{\rm Z}$ & -0.4 & 0.2 & -0.3 \\
Z+G & 18 & $L_{\rm d,BLR}^{\rm G} / L_{\rm d,line}^{\rm Z}$ & -0.4 & 0.2 & -0.5 \\
Z+X & 33 & $L_{\rm d,BLR}^{\rm X} / L_{\rm d,line}^{\rm Z}$ & -0.4 & 0.3 & -0.5 \\
\hline
M+G & 28 & $P_{\rm rl}^{\rm M}/P_{\rm fit}^{\rm G}$ & -1.0 & 0.5 & -0.9 \\
K+G & 23 & $P_{\rm rl}^{\rm K}/P_{\rm fit}^{\rm G}$ & -1.1 & 0.5 & -1.1 \\
A+G & 17 & $P_{\gamma}^{\rm A}(\Gamma_{\rm fit}^{\rm G})/P_{\rm fit}^{\rm G}$ & -0.2 & 0.4 & -0.3 \\
A+G & 17 & $P_{\gamma}^{\rm A}(\Gamma_{\rm av})/P_{\rm fit}^{\rm G}$ & -0.2 & 0.4 & -0.3 \\
P+G & 19 & $P_{\rm cs}^{\rm P}(\Gamma^{\rm G})/P_{\rm fit}^{\rm G}$ & 0.0 & 0.4 & -0.1 \\
P+M+G & 17 & $P_{\rm cs}^{\rm P}(\Gamma^{\rm G})/P_{\rm rl}^{\rm M}$ & 1.0 & 0.3 & 1.0 \\
P+K & 48 & $P_{\rm cs}^{\rm P}(\Gamma_{\rm app})/P_{\rm rl}^{\rm K}$ & 1.2 & 0.5 & 1.2 \\
\hline
G & 191 & $P_{\rm fit}^{\rm G}/\dot M_{\rm fit}^{\rm G} c^2$ & 0.0 & 0.4 & 0.0 \\
M+G & 28 & $P_{\rm rl}^{\rm M}/\dot M_{\rm fit}^{\rm G} c^2$ & -1.1 & 0.5 & -1.2 \\
K+X & 48 & $P_{\rm rl}^{\rm K}/\dot M_{\rm BLR}^{\rm X} c^2$ & -1.1 & 0.5 & -1.2 \\
A+G & 17 & $P_\gamma^{\rm A}(\Gamma^{\rm G}_{\rm fit})/\dot M_{\rm fit}^{\rm G} c^2$ & -0.3 & 0.5 & -0.3 \\
A+X & 27 & $P_\gamma^{\rm A}(\Gamma_{\rm av})/\dot M_{\rm BLR}^{\rm X} c^2$ & -0.4 & 0.6 & -0.2 \\
P+X & 15 & $P_{\rm cs}^{\rm P}(\Gamma_{\rm av})/\dot M_{\rm BLR}^{\rm X} c^2$ & 0.1 & 0.6 & 0.0 \\
P+G & 19 & $P_{\rm cs}^{\rm P}(\Gamma_{\rm fit}^G)/\dot M_{\rm fit}^{\rm G} c^2$ & 0.0 & 0.4 & 0.0 \\
Z & 54 & $P_{\rm cs}^{\rm Z}(\Gamma_{\rm app})/\dot M_{\rm line}^{\rm Z} c^2$ & -0.5 & 0.3 & -0.5 \\

%% file: jet_eff.bbl
\begin{thebibliography}{}
\makeatletter
\relax
\def\mn@urlcharsother{\let\do\@makeother \do\$\do\&\do\#\do\^\do\_\do\%\do\~}
\def\mn@doi{\begingroup\mn@urlcharsother \@ifnextchar [ {\mn@doi@}
  {\mn@doi@[]}}
\def\mn@doi@[#1]#2{\def\@tempa{#1}\ifx\@tempa\@empty \href
  {http://dx.doi.org/#2} {doi:#2}\else \href {http://dx.doi.org/#2} {#1}\fi
  \endgroup}
\def\mn@eprint#1#2{\mn@eprint@#1:#2::\@nil}
\def\mn@eprint@arXiv#1{\href {http://arxiv.org/abs/#1} {{\tt arXiv:#1}}}
\def\mn@eprint@dblp#1{\href {http://dblp.uni-trier.de/rec/bibtex/#1.xml}
  {dblp:#1}}
\def\mn@eprint@#1:#2:#3:#4\@nil{\def\@tempa {#1}\def\@tempb {#2}\def\@tempc
  {#3}\ifx \@tempc \@empty \let \@tempc \@tempb \let \@tempb \@tempa \fi \ifx
  \@tempb \@empty \def\@tempb {arXiv}\fi \@ifundefined
  {mn@eprint@\@tempb}{\@tempb:\@tempc}{\expandafter \expandafter \csname
  mn@eprint@\@tempb\endcsname \expandafter{\@tempc}}}

\bibitem[\protect\citeauthoryear{{Abdo} et~al.,}{{Abdo} et~al.}{2010}]{1FGL}
{Abdo} A.~A.,  et~al., 2010, \mn@doi [\apjs] {10.1088/0067-0049/188/2/405},
  \href {http://cdsads.u-strasbg.fr/abs/2010ApJS..188..405A} {188, 405}

\bibitem[\protect\citeauthoryear{{Arshakian}, {Le{\'o}n-Tavares},
  {B{\"o}ttcher}, {Torrealba}, {Chavushyan}, {Lister}, {Ros}  \&
  {Zensus}}{{Arshakian} et~al.}{2012}]{Arshakian2012}
{Arshakian} T.~G.,  {Le{\'o}n-Tavares} J.,  {B{\"o}ttcher} M.,  {Torrealba} J.,
   {Chavushyan} V.~H.,  {Lister} M.~L.,  {Ros} E.,   {Zensus} J.~A.,  2012,
  \mn@doi [\aap] {10.1051/0004-6361/201117140}, \href
  {http://adsabs.harvard.edu/abs/2012A%26A...537A..32A} {537, A32}

\bibitem[\protect\citeauthoryear{{Avara}, {McKinney}  \& {Reynolds}}{{Avara}
  et~al.}{2016}]{Avara2016}
{Avara} M.~J.,  {McKinney} J.~C.,   {Reynolds} C.~S.,  2016, \mn@doi [\mnras]
  {10.1093/mnras/stw1643}, \href
  {http://esoads.eso.org/abs/2016MNRAS.462..636A} {462, 636}

\bibitem[\protect\citeauthoryear{{Begelman}, {Armitage}  \&
  {Reynolds}}{{Begelman} et~al.}{2015}]{Begelman15}
{Begelman} M.~C.,  {Armitage} P.~J.,   {Reynolds} C.~S.,  2015, \mn@doi [\apj]
  {10.1088/0004-637X/809/2/118}, \href
  {http://esoads.eso.org/abs/2015ApJ...809..118B} {809, 118}

\bibitem[\protect\citeauthoryear{{Best}}{{Best}}{2009}]{Best2009}
{Best} P.~N.,  2009, \mn@doi [Astronomische Nachrichten]
  {10.1002/asna.200811152}, \href
  {http://adsabs.harvard.edu/abs/2009AN....330..184B} {330, 184}

\bibitem[\protect\citeauthoryear{{Blandford} \& {K{\"o}nigl}}{{Blandford} \&
  {K{\"o}nigl}}{1979}]{Blandford79}
{Blandford} R.~D.,  {K{\"o}nigl} A.,  1979, \mn@doi [\apj] {10.1086/157262},
  \href {http://esoads.eso.org/abs/1979ApJ...232...34B} {232, 34}

\bibitem[\protect\citeauthoryear{{Blundell} \& {Rawlings}}{{Blundell} \&
  {Rawlings}}{2000}]{Blundell00}
{Blundell} K.~M.,  {Rawlings} S.,  2000, \mn@doi [\aj] {10.1086/301254}, \href
  {http://adsabs.harvard.edu/abs/2000AJ....119.1111B} {119, 1111}

\bibitem[\protect\citeauthoryear{{Burbidge}}{{Burbidge}}{1956}]{1956Burbidge}
{Burbidge} G.~R.,  1956, \mn@doi [\apj] {10.1086/146237}, \href
  {http://adsabs.harvard.edu/abs/1956ApJ...124..416B} {124, 416}

\bibitem[\protect\citeauthoryear{{Calderone}, {Ghisellini}, {Colpi}  \&
  {Dotti}}{{Calderone} et~al.}{2013}]{Calderone2013}
{Calderone} G.,  {Ghisellini} G.,  {Colpi} M.,   {Dotti} M.,  2013, \mn@doi
  [\mnras] {10.1093/mnras/stt157}, \href
  {http://adsabs.harvard.edu/abs/2013MNRAS.431..210C} {431, 210}

\bibitem[\protect\citeauthoryear{{Celotti}, {Padovani}  \&
  {Ghisellini}}{{Celotti} et~al.}{1997}]{Celotti1997}
{Celotti} A.,  {Padovani} P.,   {Ghisellini} G.,  1997, \mn@doi [\mnras]
  {10.1093/mnras/286.2.415}, \href
  {http://adsabs.harvard.edu/abs/1997MNRAS.286..415C} {286, 415}

\bibitem[\protect\citeauthoryear{{Clausen-Brown}, {Savolainen}, {Pushkarev},
  {Kovalev}  \& {Zensus}}{{Clausen-Brown} et~al.}{2013}]{Clausen13}
{Clausen-Brown} E.,  {Savolainen} T.,  {Pushkarev} A.~B.,  {Kovalev} Y.~Y.,
  {Zensus} J.~A.,  2013, \mn@doi [\aap] {10.1051/0004-6361/201322203}, \href
  {http://esoads.eso.org/abs/2013A%26A...558A.144C} {558, A144}

\bibitem[\protect\citeauthoryear{{Croston}, {Hardcastle}, {Harris}, {Belsole},
  {Birkinshaw}  \& {Worrall}}{{Croston} et~al.}{2005}]{2005Croston}
{Croston} J.~H.,  {Hardcastle} M.~J.,  {Harris} D.~E.,  {Belsole} E.,
  {Birkinshaw} M.,   {Worrall} D.~M.,  2005, \mn@doi [\apj] {10.1086/430170},
  \href {http://adsabs.harvard.edu/abs/2005ApJ...626..733C} {626, 733}

\bibitem[\protect\citeauthoryear{{Fernandes} et~al.,}{{Fernandes}
  et~al.}{2011}]{Fernandes2011}
{Fernandes} C.~A.~C.,  et~al., 2011, \mn@doi [\mnras]
  {10.1111/j.1365-2966.2010.17820.x}, \href
  {http://adsabs.harvard.edu/abs/2011MNRAS.411.1909F} {411, 1909}

\bibitem[\protect\citeauthoryear{{Ghisellini} \& {Tavecchio}}{{Ghisellini} \&
  {Tavecchio}}{2009}]{Ghisellini09}
{Ghisellini} G.,  {Tavecchio} F.,  2009, \mn@doi [\mnras]
  {10.1111/j.1365-2966.2009.15007.x}, \href
  {http://esoads.eso.org/abs/2009MNRAS.397..985G} {397, 985}

\bibitem[\protect\citeauthoryear{{Ghisellini} \& {Tavecchio}}{{Ghisellini} \&
  {Tavecchio}}{2010}]{Ghisellini10}
{Ghisellini} G.,  {Tavecchio} F.,  2010, \mn@doi [\mnras]
  {10.1111/j.1745-3933.2010.00952.x}, \href
  {http://esoads.eso.org/abs/2010MNRAS.409L..79G} {409, L79}

\bibitem[\protect\citeauthoryear{{Ghisellini}, {Tavecchio}  \&
  {Chiaberge}}{{Ghisellini} et~al.}{2005}]{Ghisellini05}
{Ghisellini} G.,  {Tavecchio} F.,   {Chiaberge} M.,  2005, \mn@doi [\aap]
  {10.1051/0004-6361:20041404}, \href
  {http://esoads.eso.org/abs/2005A%26A...432..401G} {432, 401}

\bibitem[\protect\citeauthoryear{{Ghisellini}, {Tavecchio}, {Maraschi},
  {Celotti}  \& {Sbarrato}}{{Ghisellini} et~al.}{2014}]{Ghisellini14}
{Ghisellini} G.,  {Tavecchio} F.,  {Maraschi} L.,  {Celotti} A.,   {Sbarrato}
  T.,  2014, \mn@doi [\nat] {10.1038/nature13856}, \href
  {http://adsabs.harvard.edu/abs/2014Natur.515..376G} {515, {376 (G14)}}

\bibitem[\protect\citeauthoryear{{Godfrey} \& {Shabala}}{{Godfrey} \&
  {Shabala}}{2013}]{Godfrey13}
{Godfrey} L.~E.~H.,  {Shabala} S.~S.,  2013, \mn@doi [\apj]
  {10.1088/0004-637X/767/1/12}, \href
  {http://esoads.eso.org/abs/2013ApJ...767...12G} {767, 12}

\bibitem[\protect\citeauthoryear{{Heinz} \& {Begelman}}{{Heinz} \&
  {Begelman}}{2000}]{Heinz00}
{Heinz} S.,  {Begelman} M.~C.,  2000, \mn@doi [\apj] {10.1086/308820}, \href
  {http://esoads.eso.org/abs/2000ApJ...535..104H} {535, 104}

\bibitem[\protect\citeauthoryear{{Kharb}, {Lister}  \& {Cooper}}{{Kharb}
  et~al.}{2010}]{Kharb2010}
{Kharb} P.,  {Lister} M.~L.,   {Cooper} N.~J.,  2010, \mn@doi [\apj]
  {10.1088/0004-637X/710/1/764}, \href
  {http://adsabs.harvard.edu/abs/2010ApJ...710..764K} {710, 764}

\bibitem[\protect\citeauthoryear{{Leahy}}{{Leahy}}{1991}]{leahy91}
{Leahy} J.~P.,  1991, in {Hughes} P.~A.,  ed., Beams and Jets in Astrophysics.
  Cambridge Univ. Press, Cambridge, p.~100

\bibitem[\protect\citeauthoryear{{Ledlow} \& {Owen}}{{Ledlow} \&
  {Owen}}{1996}]{Ledlow1996}
{Ledlow} M.~J.,  {Owen} F.~N.,  1996, \mn@doi [\aj] {10.1086/117985}, \href
  {http://adsabs.harvard.edu/abs/1996AJ....112....9L} {112, 9}

\bibitem[\protect\citeauthoryear{{Levinson}}{{Levinson}}{2006}]{Levinson06}
{Levinson} A.,  2006, \mn@doi [Int. J. Mod. Phys. A]
  {10.1142/S0217751X06035063}, \href
  {http://esoads.eso.org/abs/2006IJMPA..21.6015L} {21, 6015}

\bibitem[\protect\citeauthoryear{{Lister} et~al.,}{{Lister}
  et~al.}{2009}]{Lister2009}
{Lister} M.~L.,  et~al., 2009, \mn@doi [\aj] {10.1088/0004-6256/137/3/3718},
  \href {http://adsabs.harvard.edu/abs/2009AJ....137.3718L} {137, 3718}

\bibitem[\protect\citeauthoryear{{Lobanov}}{{Lobanov}}{1998}]{Lobanov98}
{Lobanov} A.~P.,  1998, \aap, \href
  {http://esoads.eso.org/abs/1998A%26A...330...79L} {330, 79}

\bibitem[\protect\citeauthoryear{{McKinney}, {Tchekhovskoy}  \&
  {Blandford}}{{McKinney} et~al.}{2012}]{McKinney2012}
{McKinney} J.~C.,  {Tchekhovskoy} A.,   {Blandford} R.~D.,  2012, \mn@doi
  [\mnras] {10.1111/j.1365-2966.2012.21074.x}, \href
  {http://adsabs.harvard.edu/abs/2012MNRAS.423.3083M} {423, 3083}

\bibitem[\protect\citeauthoryear{{Meyer}, {Fossati}, {Georganopoulos}  \&
  {Lister}}{{Meyer} et~al.}{2011}]{Meyer2011}
{Meyer} E.~T.,  {Fossati} G.,  {Georganopoulos} M.,   {Lister} M.~L.,  2011,
  \mn@doi [\apj] {10.1088/0004-637X/740/2/98}, \href
  {http://adsabs.harvard.edu/abs/2011ApJ...740...98M} {740, 98}

\bibitem[\protect\citeauthoryear{{Narayan}, {Igumenshchev}  \&
  {Abramowicz}}{{Narayan} et~al.}{2003}]{Narayan2003}
{Narayan} R.,  {Igumenshchev} I.~V.,   {Abramowicz} M.~A.,  2003, \mn@doi
  [\pasj] {10.1093/pasj/55.6.L69}, \href
  {http://adsabs.harvard.edu/abs/2003PASJ...55L..69N} {55, L69}

\bibitem[\protect\citeauthoryear{{Nemmen}, {Georganopoulos}, {Guiriec},
  {Meyer}, {Gehrels}  \& {Sambruna}}{{Nemmen} et~al.}{2012}]{Nemmen12}
{Nemmen} R.~S.,  {Georganopoulos} M.,  {Guiriec} S.,  {Meyer} E.~T.,  {Gehrels}
  N.,   {Sambruna} R.~M.,  2012, \mn@doi [Science] {10.1126/science.1227416},
  \href {http://esoads.eso.org/abs/2012Sci...338.1445N} {338, 1445}

\bibitem[\protect\citeauthoryear{{Nolan} et~al.,}{{Nolan} et~al.}{2012}]{2FGL}
{Nolan} P.~L.,  et~al., 2012, \mn@doi [\apjs] {10.1088/0067-0049/199/2/31},
  \href {http://adsabs.harvard.edu/abs/2012ApJS..199...31N} {199, 31}

\bibitem[\protect\citeauthoryear{{Pacholczyk}}{{Pacholczyk}}{1970}]{1970Pacholczyk}
{Pacholczyk} A.~G.,  1970, {Radio astrophysics. Nonthermal processes in
  galactic and extragalactic sources}

\bibitem[\protect\citeauthoryear{{Punsly}}{{Punsly}}{2007}]{Punsly2007}
{Punsly} B.,  2007, \mn@doi [\mnras] {10.1111/j.1745-3933.2006.00254.x}, \href
  {http://adsabs.harvard.edu/abs/2007MNRAS.374L..10P} {374, L10}

\bibitem[\protect\citeauthoryear{{Punsly}}{{Punsly}}{2011}]{Punsly2011}
{Punsly} B.,  2011, \mn@doi [\apjl] {10.1088/2041-8205/728/1/L17}, \href
  {http://adsabs.harvard.edu/abs/2011ApJ...728L..17P} {728, L17}

\bibitem[\protect\citeauthoryear{{Punsly} \& {Zhang}}{{Punsly} \&
  {Zhang}}{2011}]{PunslyZhang2011}
{Punsly} B.,  {Zhang} S.,  2011, \mn@doi [\mnras]
  {10.1111/j.1745-3933.2011.01019.x}, \href
  {http://adsabs.harvard.edu/abs/2011MNRAS.412L.123P} {412, L123}

\bibitem[\protect\citeauthoryear{{Pushkarev}, {Kovalev}, {Lister}  \&
  {Savolainen}}{{Pushkarev} et~al.}{2009}]{Pushkarev09}
{Pushkarev} A.~B.,  {Kovalev} Y.~Y.,  {Lister} M.~L.,   {Savolainen} T.,  2009,
  \mn@doi [\aap] {10.1051/0004-6361/200913422}, \href
  {http://esoads.eso.org/abs/2009A%26A...507L..33P} {507, L33}

\bibitem[\protect\citeauthoryear{{Pushkarev}, {Hovatta}, {Kovalev}, {Lister},
  {Lobanov}, {Savolainen}  \& {Zensus}}{{Pushkarev}
  et~al.}{2012}]{Pushkarev2012}
{Pushkarev} A.~B.,  {Hovatta} T.,  {Kovalev} Y.~Y.,  {Lister} M.~L.,  {Lobanov}
  A.~P.,  {Savolainen} T.,   {Zensus} J.~A.,  2012, \mn@doi [\aap]
  {10.1051/0004-6361/201219173}, \href
  {http://adsabs.harvard.edu/abs/2012A%26A...545A.113P} {545, A113}

\bibitem[\protect\citeauthoryear{{Rawlings} \& {Saunders}}{{Rawlings} \&
  {Saunders}}{1991}]{Rawlings1991}
{Rawlings} S.,  {Saunders} R.,  1991, \mn@doi [\nat] {10.1038/349138a0}, \href
  {http://adsabs.harvard.edu/abs/1991Natur.349..138R} {349, 138}

\bibitem[\protect\citeauthoryear{{Shabala} \& {Godfrey}}{{Shabala} \&
  {Godfrey}}{2013}]{Shabala13}
{Shabala} S.~S.,  {Godfrey} L.~E.~H.,  2013, \mn@doi [\apj]
  {10.1088/0004-637X/769/2/129}, \href
  {http://esoads.eso.org/abs/2013ApJ...769..129S} {769, 129}

\bibitem[\protect\citeauthoryear{{Shabala}, {Santoso}  \& {Godfrey}}{{Shabala}
  et~al.}{2012}]{Shabala12}
{Shabala} S.~S.,  {Santoso} J.~S.,   {Godfrey} L.~E.~H.,  2012, \mn@doi [\apj]
  {10.1088/0004-637X/756/2/161}, \href
  {http://esoads.eso.org/abs/2012ApJ...756..161S} {756, 161}

\bibitem[\protect\citeauthoryear{{Sikora}}{{Sikora}}{2016}]{sikora16b}
{Sikora} M.,  2016, in {G\'{o}mez} J.~L.,  {Marscher} A.~P.,   {Jorstad} S.~G.,
   eds, Blazars through Sharp Multi-Wavelength Eyes. Galaxies, in press

\bibitem[\protect\citeauthoryear{{Sikora}, {Stasi{\'n}ska},
  {Kozie{\l}-Wierzbowska}, {Madejski}  \& {Asari}}{{Sikora}
  et~al.}{2013}]{Sikora2013}
{Sikora} M.,  {Stasi{\'n}ska} G.,  {Kozie{\l}-Wierzbowska} D.,  {Madejski}
  G.~M.,   {Asari} N.~V.,  2013, \mn@doi [\apj] {10.1088/0004-637X/765/1/62},
  \href {http://adsabs.harvard.edu/abs/2013ApJ...765...62S} {765, 62}

\bibitem[\protect\citeauthoryear{{Sikora}, {Rutkowski}  \& {Begelman}}{{Sikora}
  et~al.}{2016}]{Sikora2016}
{Sikora} M.,  {Rutkowski} M.,   {Begelman} M.~C.,  2016, \mn@doi [\mnras]
  {10.1093/mnras/stw107}, \href
  {http://adsabs.harvard.edu/abs/2016MNRAS.457.1352S} {457, 1352}

\bibitem[\protect\citeauthoryear{{Tchekhovskoy}, {McKinney}  \&
  {Narayan}}{{Tchekhovskoy} et~al.}{2009}]{Tchekhovskoy2009}
{Tchekhovskoy} A.,  {McKinney} J.~C.,   {Narayan} R.,  2009, \mn@doi [\apj]
  {10.1088/0004-637X/699/2/1789}, \href
  {http://adsabs.harvard.edu/abs/2009ApJ...699.1789T} {699, 1789}

\bibitem[\protect\citeauthoryear{{Tchekhovskoy}, {Narayan}  \&
  {McKinney}}{{Tchekhovskoy} et~al.}{2011}]{Tchekhovskoy2011}
{Tchekhovskoy} A.,  {Narayan} R.,   {McKinney} J.~C.,  2011, \mn@doi [\mnras]
  {10.1111/j.1745-3933.2011.01147.x}, \href
  {http://adsabs.harvard.edu/abs/2011MNRAS.418L..79T} {418, L79}

\bibitem[\protect\citeauthoryear{\VAN{Velzen}{van}{van}~Velzen \&
  {Falcke}}{\VAN{Velzen}{van}{van}~Velzen \& {Falcke}}{2013}]{vanVelzen2013}
\VAN{Velzen}{van}{van}~Velzen S.,  {Falcke} H.,  2013, \mn@doi [\aap]
  {10.1051/0004-6361/201322127}, \href
  {http://adsabs.harvard.edu/abs/2013A%26A...557L...7V} {557, L7}

\bibitem[\protect\citeauthoryear{{Volonteri}, {Sikora}, {Lasota}  \&
  {Merloni}}{{Volonteri} et~al.}{2013}]{Volonteri2013}
{Volonteri} M.,  {Sikora} M.,  {Lasota} J.-P.,   {Merloni} A.,  2013, \mn@doi
  [\apj] {10.1088/0004-637X/775/2/94}, \href
  {http://adsabs.harvard.edu/abs/2013ApJ...775...94V} {775, 94}

\bibitem[\protect\citeauthoryear{{Willott}, {Rawlings}, {Blundell}  \&
  {Lacy}}{{Willott} et~al.}{1999}]{Willott1999}
{Willott} C.~J.,  {Rawlings} S.,  {Blundell} K.~M.,   {Lacy} M.,  1999, \mn@doi
  [\mnras] {10.1046/j.1365-8711.1999.02907.x}, \href
  {http://adsabs.harvard.edu/abs/1999MNRAS.309.1017W} {309, 1017}

\bibitem[\protect\citeauthoryear{{Xiong} \& {Zhang}}{{Xiong} \&
  {Zhang}}{2014}]{XiongZhang2014}
{Xiong} D.~R.,  {Zhang} X.,  2014, \mn@doi [\mnras] {10.1093/mnras/stu755},
  \href {http://adsabs.harvard.edu/abs/2014MNRAS.441.3375X} {441, 3375}

\bibitem[\protect\citeauthoryear{{Zamaninasab}, {Clausen-Brown}, {Savolainen}
  \& {Tchekhovskoy}}{{Zamaninasab} et~al.}{2014}]{Zamaninasab14}
{Zamaninasab} M.,  {Clausen-Brown} E.,  {Savolainen} T.,   {Tchekhovskoy} A.,
  2014, \mn@doi [\nat] {10.1038/nature13399}, \href
  {http://adsabs.harvard.edu/abs/2014Natur.510..126Z} {510, {126 (Z14)}}

\bibitem[\protect\citeauthoryear{{Zdziarski}}{{Zdziarski}}{2014}]{2014Zdziarski}
{Zdziarski} A.~A.,  2014, \mn@doi [\mnras] {10.1093/mnras/stu1835}, \href
  {http://adsabs.harvard.edu/abs/2014MNRAS.445.1321Z} {445, 1321}

\bibitem[\protect\citeauthoryear{Zdziarski, Sikora, Pjanka  \&
  Tchekhovskoy}{Zdziarski et~al.}{2015}]{Zdziarski15}
Zdziarski A.~A.,  Sikora M.,  Pjanka P.,   Tchekhovskoy A.,  2015, \mn@doi
  [\mnras] {10.1093/mnras/stv986}, 451, {927 (Z15)}

\makeatother
\end{thebibliography}
